\documentclass[11pt]{article}
\usepackage{times}
\usepackage{geometry}
\usepackage[utf8]{inputenc}
\usepackage{enumitem,amssymb}
\usepackage{graphicx}
\usepackage{fancyhdr}
\usepackage{aas_macros}
\usepackage{mdframed} 

\usepackage[authoryear]{natbib}
\setcitestyle{authoryear,open={(},close={)}}

\mdfdefinestyle{theoremstyle}{
innertopmargin=\topskip,}
\mdtheorem[style=theoremstyle]{lrptextbox}{}

\begin{document}
\title{The Next Generation Very Large Array}
\author{James Di Francesco (NRC), Dean Chalmers (NRC), Nolan Denman (NRAO) \\
Laura Fissel (Queen's), Rachel Friesen (Toronto), Bryan Gaensler (Toronto), \\
Julie Hlavacek-Larrondo (Montreal), Helen Kirk (NRC), Brenda Matthews (NRC), \\
Christopher O'Dea (Manitoba), Tim Robishaw (NRC), Erik Rosolowsky (Alberta), \\
Michael Rupen (NRC), Sarah Sadavoy (Queen's), Samar Safi-Harb (Manitoba) \\
Greg Sivakoff (Alberta), Mehrnoosh Tahani (NRC), Nienke van der Marel (NRC), \\
Jacob White (Konkoly Obs.), and Christine Wilson (McMaster)}
\maketitle

\begin{abstract}
The next generation Very Large Array (ngVLA) is a transformational radio observatory being designed by the U.S. National Radio Astronomy Observatory (NRAO). It will provide order of magnitude improvements in sensitivity, resolution, and uv coverage over the current Jansky Very Large Array (VLA) at $\sim$1.2-50 GHz and extend the frequency range up to 70-115 GHz. This document is a white paper written by members of the Canadian community for the 2020 Long Range Plan panel, which will be making recommendations on Canada's future directions in astronomy. Since Canadians have been historically major users of the VLA and have been valued partners with NRAO for ALMA, Canada's participation in ngVLA is welcome. Canadians have been actually involved in ngVLA discussions for the past five years, and have played leadership roles in the ngVLA Science and Technical Advisory Councils. Canadian technologies are also very attractive for the ngVLA, in particular our designs for radio antennas, receivers, correlates, and data archives, and our industrial capacities to realize them. Indeed, the Canadian designs for the ngVLA antennas and correlator/beamformer are presently the baseline models for the project. Given the size of Canada's radio community and earlier use of the VLA (and ALMA), we recommend Canadian participation in the ngVLA at the 7\% level. Such participation would be significant enough to allow Canadian leadership in gVLA's construction and usage. Canada's participation in ngVLA should not preclude its participation in SKA; access to both facilities is necessary to meet Canada's radio astronomy needs. Indeed, ngVLA will fill the gap between those radio frequencies observable with the SKA and ALMA at high sensitivities and resolutions. Canada's partnership in ngVLA will give it access to cutting-edge facilities together covering approximately three orders of magnitude in frequency.
\end{abstract}


%


~

\section{The Next Generation Very Large Array (ngVLA)}
The ngVLA will be a transformational radio observatory that will probe the formation of terrestrial zone planets, the astronomical origins of biological chemicals, the evolution of galaxies across cosmic time, and the environs of black holes.  It will replace the existing, but aging, Jansky Very Large Array (VLA) by offering order of magnitude improvements in sensitivity, resolution, and instantaneous $uv$ coverage at 1-50 GHz and extending the observational frequency range into the millimetre regime (to 115 GHz).  The ngVLA project is being led by the U.S. National Radio Astronomy Observatory (NRAO), with the goal of international participation at the 25\% level.  Notably, Canadians have been involved in the definition of ngVLA from the very earliest discussions.

The ngVLA concept includes 263 radio antennas centered at the current VLA site on the Plains of San Agustin in central New Mexico, U.S.A.  The observatory will actually consist of three separate arrays working in parallel: 

\begin{itemize}
    \item a {\it Main Array} (MA) of 214 $\times$ 18-m antennas clustered mostly in New Mexico (see Fig. 1), but with several located as far away as west Texas, southern Arizona, and northern Mexico, providing 10-1000 km baselines;
	\item a {\it Short Baseline Array} (SBA) of 19 $\times$ 6-m antennas located at the current VLA site (+ 4 inner MA antennas with total-power capabilities), allowing low surface brightness imaging, and;
	\item a {\it Long Baseline Array} (LBA) of 30 $\times$ 18-m antennas located in 10 clusters across the U.S. from Hawaii to the U.S. Virgin Islands but also western Canada, allowing very high angular resolution imaging. 
\end{itemize}

\noindent
Signals received from all ngVLA antennas will be transported via optical fiber to a central correlator also located at the current VLA site in New Mexico.  

The ngVLA is presently under consideration by the Astro2020 decadal survey of astronomy and astrophysics currently underway in the U.S.A.  It is targeted to begin early operations in the late 2020s and full operations in the mid-2030s.  
The ngVLA is an excellent scientific and technological opportunity for Canada in the coming decades. 

~

\begin{figure}[ht]
 \centerline{\includegraphics[width=0.9\textwidth]{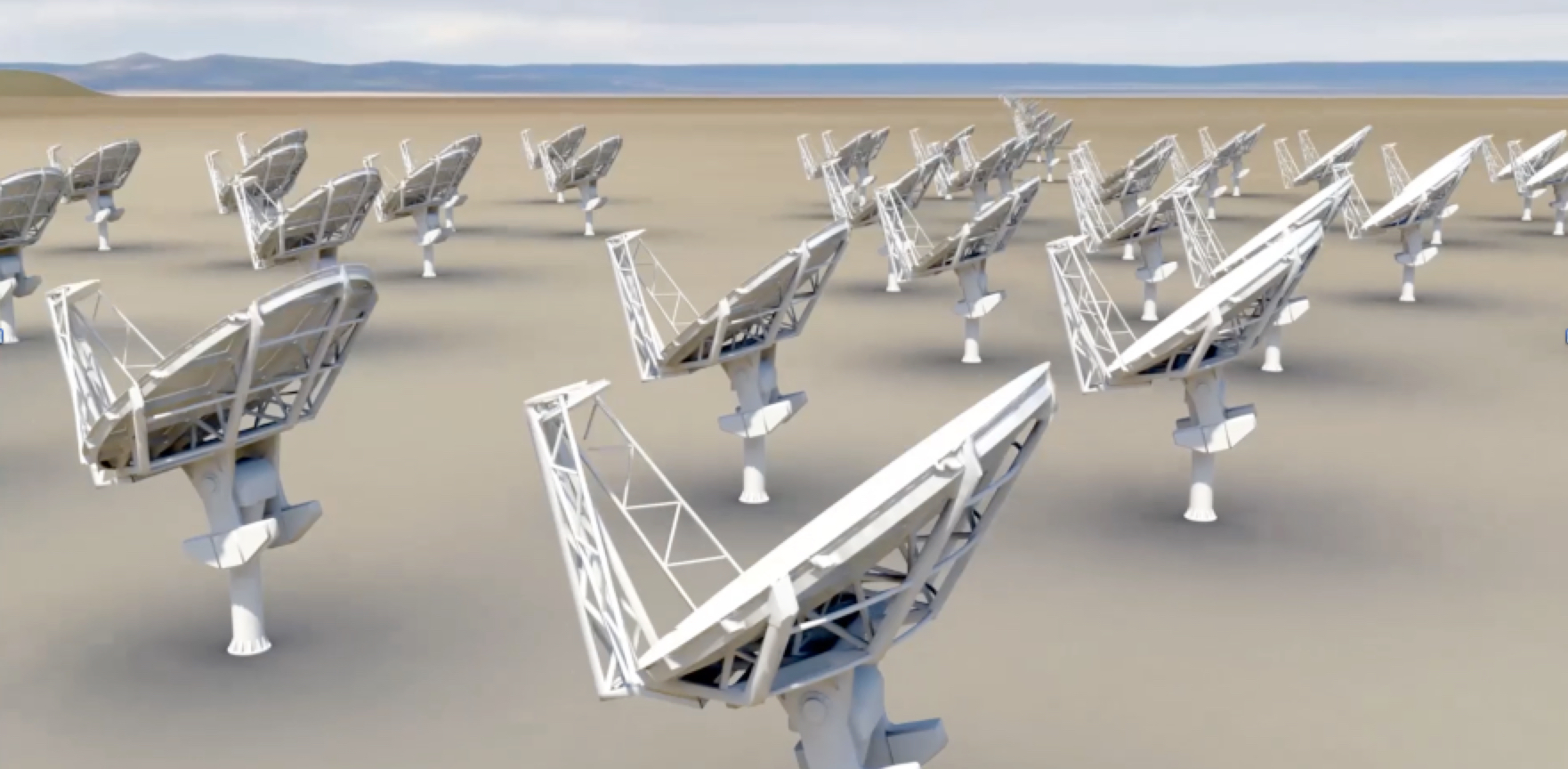}}
 \caption{\label{fig:ngVLA} Render of the ngVLA concept showing its central antennas in New Mexico.}
\end{figure}

\section{Science Drivers}

The ngVLA design has been shaped by nearly five years of extensive discussion within the international astronomical community about its radio astronomy needs in the forthcoming era of the James Webb Space Telescope (JWST), the Extremely Large Telescopes (ELTs), the Atacama Large Millimeter/submillimeter Array (ALMA), the Square Kilometer Array (SKA), and now multi-messenger astrophysics.   Specifically,  ngVLA will complementarily bridge radio frequencies  between those observable with SKA and ALMA at high sensitivity and resolution.

After successive conferences about the future of radio astronomy starting in early 2015, NRAO struck the ngVLA Science Advisory Council (SAC) and the Technical Advisory Council (TAC), each with Canadian members, to guide evolution of the ngVLA concept.  Subsequent SAC and TAC discussions in 2017 led to the development of $\sim$80 mature community-driven science use cases for a large mm/cm interferometer that span a very wide range of astrophysics.  Building the scientific case further, the {\it ngVLA Science Book} was published in early 2019 with 88 science chapters from 285 unique authors on topics touching on almost all aspects of astrophysics.

A major result of the community consultation process was the identification of five Key Science Goals (KSGs) for an ngVLA, i.e., high-impact projects that can be addressed uniquely by a next generation large mm/cm interferometer.  We describe these KSGs here.  (Please see the {\it ngVLA Science Book} for more detail and references.)

\subsection{KSG1: Unveiling the Formation of Solar System Analogues on Terrestrial Scales}

ALMA has driven forward our understanding of planet formation, with stunning detections at its highest resolutions of gap-like structures in numerous circumstellar disks that have been arguably cleared by young planets (see Fig. 2, far left).  Though ALMA will make significant further inroads in this field for years to come by observing more disks, it is unable to resolve the so-called “terrestrial zone” of $\sim$1 au scales for disks in the nearest star-forming regions at $\sim$125 pc. For context, ALMA at its highest resolution has only resolved the cleared terrestrial zone of the TW Hya disk, which at 50 pc makes it the closest young disk to the Sun. 

The ngVLA will observe hundreds of protoplanetary disks at $\sim$5 mas resolution, allowing the terrestrial zones of disks to be well characterized in the nearest star-forming regions, e.g., 0.6 au resolution at d = 125 pc.  Fig. 2 (far left) also shows how only 5 hr of integration with the ngVLA will reveal gaps and rings produced by planets in the terrestrial zone, as well as disks around protoplanets and other disk features shaped by young gas giants (e.g., crescents).  These data will reveal the initial mass function of planets down to 5-10 M$_{Earth}$ and the planet birth radius distribution from the inner edges of disks to tens of au (see Fig. 2, middle left). Furthermore, ngVLA is the only facility planned that will reveal the temporal evolution of inner disk features via multi-epoch observations.

\subsection{KSG2: Probing the Initial Conditions for Planetary Systems and Life with Astrochemistry}

Emission from numerous simple interstellar molecules has been detected over the past 50 years, clarifying the chemical processes in play in clouds and disks.   Prebiotic molecules, however, have been harder to detect because their innate complexities allow numerous internal modes that distribute excitation over many transitions.  ALMA’s high sensitivity in the submillimetre range has proven to be beneficial but line emission there can be extremely crowded to the point of confusion, making weaker emission from complex molecules hard to discern.  

The ngVLA will have extremely high sensitivities to line emission at frequency ranges that are far less crowded, i.e., those lower than ALMA can observe.  Its high sensitivity and resolution will enable the chemical composition of planet-forming environments, especially the disk mid-planes, to be probed both spatially and temporally as never before.  For example, snowlines of critical species like ammonia will be particularly well probed by the ngVLA.  In addition, the ngVLA will have the sensitivity and frequency coverage to detect for the first time the interstellar signatures of amino acids such as glycine and alanine, as well as their precursors and other key prebiotic molecules, providing direct insight into the chemical building blocks of life and how life arose on Earth.    

\subsection{KSG3: Charting the Assembly, Structure, and Evolution of Galaxies from the First Billion Years to the Present}

Revealing the star-formation history of galaxies from the Epoch of Reionization (EoR) forward is a major goal of JWST and the ELTs.  Complementary to such probes will be investigations of the evolution of gas within galaxies over the same important range of redshift.  Indeed, galactic gas content data are needed to understand the evolution of the reservoirs for star formation across cosmic time, informing how gas accretes from the intergalactic medium (IGM) and affects local star formation efficiencies.  ALMA is providing key insights into the high excitation regions of molecular clouds but by virtue of its frequency range it cannot detect the highly redshifted lower-{\it J} transitions of CO that are the best tracers of molecular mass content in younger galaxies.

The ngVLA will measure routinely the gas content of “main-sequence” galaxies out to the EoR and beyond by tracing redshifted low-{\it J} CO emission (see Fig. 2, middle right). Thousands of galaxies out to {\it z} = 8 will be observed in large volume surveys within the first 10 years of ngVLA operations, enabling precision measurements of the evolution of molecular gas over cosmic time (see Fig. 2, far right).  In addition, the ngVLA will also sensitively detect HI emission at arcsecond resolution, probing the accretion of atomic gas onto and through galaxies, the conversion between diffuse atomic gas into the molecular clouds that fuel star formation, and the cold outflow of atomic gas back into the IGM. 

\subsection{KSG4: Using Galactic Centre Pulsars to Make a Fundamental Test of General Relativity}

Our Galactic Centre contains the closest example of a supermassive black hole, i.e., Sgr A*.  Indeed, infrared imaging with adaptive optics over the past 20 years has revealed the motions of stars orbiting Sgr A* and thus enabled the accurate measurement of the black hole’s mass to 4 billion M$_{\odot}$.  Our relative proximity to the extreme gravitational field surrounding Sgr A* thus makes the Galactic Centre a very special environment where General Relativity can be tested to new ways.  For example, the periods of pulsars can be traced to extremely high precision, providing stable clocks that would be affected by the variations in the spacetime potential surrounding Sgr A*.  Far fewer pulsars than expected have been detected in the Galactic Centre, however, possibly due to high levels of radio scattering through the Galactic Plane.  

The ngVLA will be able to examine the Galactic Centre to very high sensitivity at frequencies where radio scattering will be diminished, e.g., $>$10 GHz, to provide the best probes yet for the expected pulsar population around Sgr A*.  The ngVLA will then subsequently monitor these pulsars to test General Relativity to unprecedented accuracy.  Even not detecting new pulsars in the Galactic Centre would be interesting.  Indeed, the recent discovery of a radio magnetar in the Galactic Centre has called into question radio scattering as the reason behind the deficit of pulsars in the region.  A definitive result that the number of pulsars in the Galactic Centre region is smaller than expected could have profound implications about star formation and stellar evolution in that environment.

\subsection{KSG5: Understanding the Formation and Evolution of Stellar and Supermassive Black Holes and Compact Objects in the Era of Multi-messenger Astronomy}

The detection of mergers of compact extreme objects like neutron stars and black holes via gravitational waves in the past few years has ushered in a new era of multi-messenger astronomy.  In the coming decades, upgraded or new gravitational wave observatories (e.g., Advanced LIGO, VIRGO, LISA) will enable detections of compact object mergers out to 200 Mpc.  Characterizations of the electromagnetic counterparts of the mergers, however, are important as they provide key contexts for these very energetic events.

With its high sensitivity, astrometric resolutions, and rapid response capabilities, the ngVLA will be the premier electromagnetic probe of black hole formation and growth.  In our Galaxy, the ngVLA will expand the number of known black hole X-ray binaries by an order of magnitude.  The precise proper motions measured by the ngVLA from these objects (e.g., 20 km/s at 5 kpc) will be used to trace their natal kick velocities, distances, and masses.  Beyond the Galaxy, the ngVLA will trace the innermost regions of globular clusters and dwarf galaxies out to 20~Mpc for the elusive population of intermediate-mass black holes.  Furthermore, the ngVLA will probe the phases of supermassive black hole mergers out to 200 Mpc by measuring their binary orbital motions at sub-pc scales.  Finally, the highest resolutions provided by the continental-scale baselines of the ngVLA will probe the mechanisms behind the acceleration of high-energy neutrinos from AGN jets.

\begin{figure}[ht]
 \centerline{\includegraphics[width=1.0\textwidth]{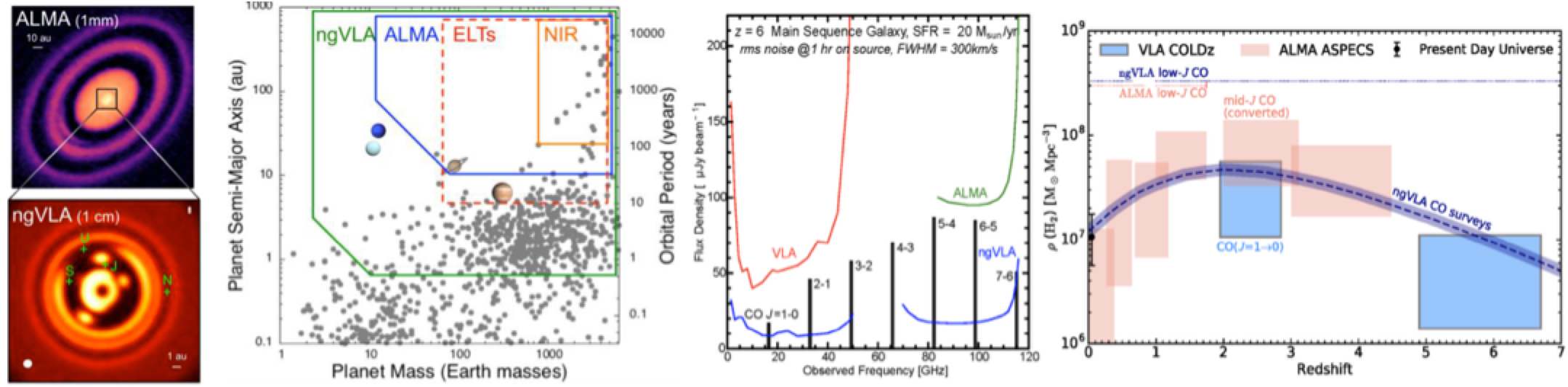}}
 \caption{\label{fig:ksgs} {\it Far left}: ({\it above}) Actual 1 mm ALMA observations of a protoplanetary disk at 0.04$^{\prime\prime}$ resolution, and ({\it below}) simulated 1 cm ngVLA observations of a terrestrial zone at 0.01$^{\prime\prime}$ resolution.  {\it Middle left}: Distribution of exoplanet masses and orbits with regions probed by various facilities highlighted.  {\it Middle right}: Sensitivity of ngVLA to detect CO emission from a $z$ = 6 galaxy, vs. VLA and ALMA.  {\it Far right}: Improvement of knowledge of cold gas density over cosmic time expected by ngVLA vs. current VLA ({\it blue}) and ALMA ({\it red}) programs.}
\end{figure}

\subsection{Other Science}
The five Key Science Goals highlight the transformational nature of the ngVLA. They have been used to inform the operational specifications of the observatory, e.g., its frequency range, its sensitivities, its angular resolutions, and the largest recoverable scales (see next section).  These specifications, however, also provide considerable versatility and enable a wide range of other exciting astrophysics, as described in the {\it ngVLA Science Book} and in numerous conferences over the past few years.  

Indeed, ngVLA will be very important for future projects led by Canadians, e.g., KSG1: Matthews or Dong, KSG3: Rosolowsky, Wilson, or Irwin, and KSG5: Hlavacek-Larrondo.  Beyond the Key Science Goals, however, exciting possibilities include i) access to high-sensitivity data of low surface brightness ammonia emission from dense gas of star-forming molecular clouds via the SBA+MA, as evidenced by recent work by Di Francesco, Kirk, or Friesen; ii) access to high-resolution polarization data via the MA, as evidenced by recent work by Sadavoy, Fissel, Robishaw, or Bastien, and; iii) access to extremely high-resolution observations via the LBA, as evidenced by recent work on many topics by Bartel, Bietenholz, Heinke, Pen, Rupen, or Sivakoff.

\section{ngVLA Specifications}

Here we provide a summary of the key performance requirements expected of the ngVLA:

\begin{itemize}
\item {\it Frequency Ranges}: The ngVLA will observe any frequency between $\sim$1.2 GHz and 115 GHz, except in the 50-70 GHz range where the atmosphere is opaque.  The low end will allow observations of HI 21-cm emission and pulsars (KSGs 3 and 4).  The high end, at much higher frequencies that the VLA’s current maximum frequency of 50~GHz, will allow observations of the low-{\it J} molecular transitions (e.g., CO 1-0) and thermal dust emission (KSG 1).  All antennas will observe this frequency range.

\item {\it Continuum Sensitivities}: The MA will achieve point-source sensitivities of 0.07~$\mu$Jy/beam at 30~GHz in 8~hours of integration or 0.5~$\mu$Jy/beam at 100~GHz in 2~hours needed to detect continuum emission from protoplanetary disks (KSG 1).  The added LBA will achieve the sensitivities of 0.23~$\mu$Jy/beam at 10~GHz in 1~hour of integration needed to identify the origins of gravitational wave events out to 200~Mpc (KSG 5).  

\item {\it Spectral sensitivities}: The MA+SBA will achieve the sensitivities of 30 $\mu$Jy/beam/km/s in 10 hours needed for astrochemistry studies (KSG 2) and blind surveys of gas from high-redshift galaxies (KSG 3).  They will also achieve the sensitivities of 1-750 mK at 5-0.1$^{\prime\prime}$ angular resolution and 1-20 km/s spectral resolution in 10 hours needed for studies of CO (and other molecules) from galaxies across cosmic time (KSG 3) 

\item {\it Angular Resolutions}: The MA's $\sim$0.01-1000 km baselines will give the angular resolutions of $\sim$5~mas at 30 GHz and 100 GHz needed to resolve the $\sim$1 au terrestrial zone of protoplanetary disks out to the nearest star-forming regions (125 pc; KSG 1).  Note that such resolution will be comparable to or better than those achievable from the ELTs.  The LBA's $\sim$8,800 km maximum baseline will give the angular resolutions of 0.6 mas at 10 GHz needed to measure the proper motions of gravitational wave events out to 200 Mpc (KSG 5).

\item {\it Largest Recoverable Scales}: the SBA will recover emission on angular scales of $>$20$^{\prime\prime}$ x (100 GHz/$\nu$) that are needed to detect the extended emission of molecular gas clouds in our own Galaxy and in nearby galaxies (KSG 3).

\item {\it Additional requirements}: For pulsar observations, the ngVLA will allow $>$10 formed beams spread over 1-10 subarrays that will be fed to a centralized pulsar search/timing engine (KSG 4).  For polarization observations, all four Stokes parameters will be recoverable.
\end{itemize}

\section{Operations}

The ngVLA will leverage NRAO’s considerable previous expertise constructing and operating the VLA, the Very Long Baseline Array (VLBA), the Green Bank Telescope (GBT), and ALMA.  It will be a PI proposal-driven observatory where the most compelling science of the day from the community will be pursued.  Some moderate fraction of time is expected to be devoted to larger-scale ``Legacy" projects.  Proposal calls will be sent out annually.  

Science operations, support staff, software development and maintenance, and administration services will be situated at an ngVLA Science Operations Centre and Data Centre in an accessible metropolitan location that is TBD.  Users will interact with staff at that location for assistance with proposal preparation, project execution, and data delivery.  To make the observatory as accessible as possible to the astronomical community, the ngVLA will follow the ALMA/VLA model and provide its users with “science-ready data products,” i.e., images and image cubes produced via reliable pipeline processing of visibilities, in a timely manner.  Both raw visibilities and calibration tables will be stored in the ngVLA Data Archive.

An ngVLA Antenna Operations Centre will be situated in Socorro, NM with further repair facilities at the nearby array centre.  More distant antennas will have on-site storage and annual maintenance visits.  All antennas will be fixed at their locations, and hence ngVLA will not have multiple configurations.  The ngVLA, however, will allow simultaneous observing of multiple sub-arrays in case only specific ranges of $uv$ coverage are desired.

\section{Project Costs}

NRAO has utilized extensive formal project management approaches, including independent auditing, to determine reasonable estimates of the costs of ngVLA construction and operation.  They are based on flowdown from the {\it ngVLA Reference Design}, a $\sim$1500-page suite of detailed documentation.  Table 1 shows the cost estimates in FY2018 dollars provided in the ngVLA APC white paper submitted by the project to the Astro2020 decadal survey panel, which includes cost estimates for all project phases from design/development though operations and eventual divestment.  These estimates are reported at a 70\% confidence level.

Total construction costs over FY25-34 are US\$2.25B and total operations costs over FY28-54 are US\$2.13B in FY2018 dollars.  NRAO is aiming for 25\% international participation for the design, construction and operations phases.  Operations costs include funds for general maintenance and upkeep of ngVLA equipment and infrastructure (including staff) and, following the ALMA model, funds for the continuous development of new ngVLA capabilities, archives, or technologies by the partner communities.

\begin{figure}[ht]
 \centerline{\includegraphics[width=0.9\textwidth]{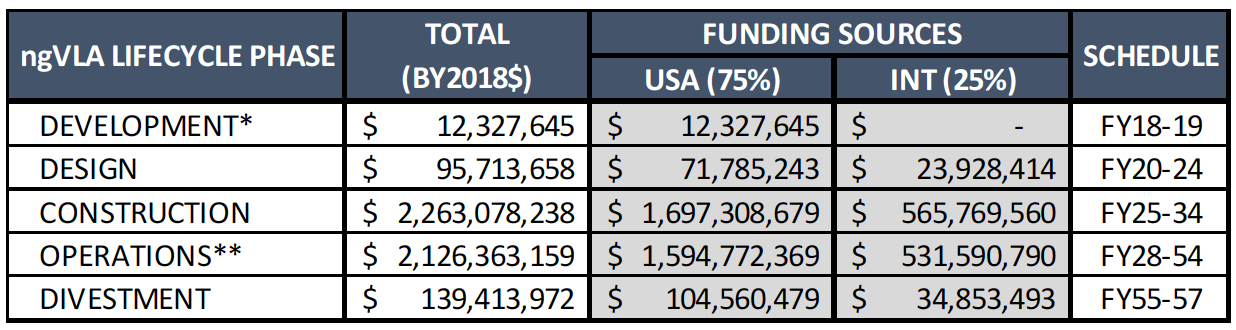}}
 \caption{\label{fig:tab1} 
 Anticipated integrated costs of ngVLA over its five lifecycle phases in FY2018 USD. *Does not include in-kind contributions from NRC.  **Annual operations estimate in FY2018 USD is \$92.9M/yr.}
\end{figure}

\section{Partnerships and Canadian Leadership}

Canada is an obvious partner for ngVLA.  Fifteen years ago, Canada joined ALMA in partnership with the U.S., providing a range of valued contributions during ALMA’s construction stage (e.g., the Band 3 receivers).  For operations, Canada continues to partner with the U.S. within the North American (NA) regional partnership, contributing 7.25\% of the NA support of ALMA.  As noted in the LRP2020 GAC and ALMA2030 White Papers, Canada has received a fair return for its contributions in observing time and development funds.  Also, Canadian use of the current VLA has been historically quite vigorous.  For example, Dennis Crabtree reports in his 2019 survey of international observatory performance that 15\% of all VLA papers have had at least one Canadian co-author with 5\% having a Canadian lead.

Canadians have been involved in ngVLA discussions with NRAO from the very start.  For example, Canadians have been active members on the Executive of the ngVLA Science Advisory Council since its inception, and have led the organization of science use case studies and ngVLA-related science meetings over the past few years.  Also, Canadians have been active members of the ngVLA Technical Advisory Council, and the Canadian antenna and beamformer/correlator designs are at present the baseline model for ngVLA.  The ngVLA project has been widely promoted within Canada, with presentations at recent CASCA and Canadian Radio Futures meetings.  The ngVLA-Canada e-mail list, with which ngVLA news is disseminated out to our community, currently has over 40 members.

Canada’s longstanding engagement in ngVLA could translate to significant opportunities for Canadian contributions to ngVLA construction.  Leveraging the NRC baseline design for ngVLA’s antennas, Canada could provide important dish technology to the project.  As antennas are the largest component of ngVLA construction costs, it is critical that the project utilize low-cost, high-performance antennas.  Though building all 263 ngVLA antennas is likely too large a contribution for Canada to make on its own, licensing its novel antenna technology to others could be itself a key contribution.   More to scale, Canada could easily provide the 19 $\times$ 6-m antennas of the SBA, based on its own designs, as a contribution.  In addition, Canada’s expertise in correlator technology cannot be understated.  Leveraging our involvement in designing SKA’s Central Signal Processor and VLA’s current WIDAR correlator, Canada could design or provide ngVLA with a state-of-the-art correlator using its new TALON architecture.  Indeed, NRAO has begun to look into unifying the designs for an upgraded ALMA correlator and the ngVLA correlator, and Canada is well placed to contribute to both activities.  Finally, Canada could provide important receiver contributions to ngVLA.  We have continued to research receiver technology after our successful run of providing ALMA with its Band 3 receivers, and contributing to its new Band 1 receivers as well.  Note that Bands 1 and 3 cover most of the high frequency range of ngVLA, i.e., 35-50 GHz and 86-115 GHz, respectively.  More recent Canadian research into SKA receivers could itself be leveraged into the most modern, high sensitivity, dual-polarization receivers possible for ngVLA.   It is important to highlight that Canadian industry will benefit substantially from Canadian participation in ngVLA.   For example, the TeraXion and Nanowave companies benefited during ALMA construction by providing key components to ALMA’s LO system and Band 3 receivers, respectively, following NRC’s policy of engaging Canadian industry with large astronomy projects.

In terms of operations, Canada could provide future in-kind support services to ngVLA in a manner similar to the effective distributed ALMA support partnership NRC has with NRAO within the North American ALMA Science Centre.  In addition, Canadians could provide significant contributions in operations to the future ngVLA Data Archive.  As with many observatories, the archive will be critical for disseminating data to its current and future users.  At present, CIRADA is a large university-NRC collaboration funded by CFI to provide archival access to data produced by the Very Large Array Sky Survey (VLASS) project and ASKAP.  Canada has also begun discussing the requirements for its SKA Regional Data Centre, allowing us access to SKA data services. Such investments could be leveraged in the future to provide ngVLA with high-quality data services at relatively low development cost.

Finally, the ngVLA baseline plan includes a set of three LBA antennas in Canada itself, i.e, at DRAO near Penticton, BC.  Given the relative proximity of the nearest station in Brewster, WA, the DRAO location will provide more accurate calibration to the LBA.

In summary, Canada is poised to provide significant leadership within the ngVLA project.  Indeed, Canada is not the only potential international partner in ngVLA.  So far, Japan, Taiwan, Mexico, Thailand, and Germany have also expressed interest in the project.   In May 2019, NRAO hosted an international workshop with Canada and other potential partners to discuss possible ngVLA contributions.  No agreements for official participation in ngVLA, however, have been signed by any partner as of September 2019.

\section{Scientific and Programmatic Risks}

\subsection{Funding}

As the ngVLA is being led by the U.S.A., its success is highly dependent on the project attaining considerable funding from the U.S. National Science Foundation (NSF).  The project has received from NSF US\$11M in late 2017 and US\$4M in late 2019 to fund design and development activities in advance of the Astro2020 U.S. decadal survey.  These activities will also enable preparation of a robust proposal for ngVLA construction funds that will be submitted in FY21 to NSF’s Major Research Equipment and Facilities Construction (MREFC) program. 

Enthusiastic support from the Astro2020 panel is critical for the ngVLA to attract significant investment from the NSF.  So far, the U.S. astronomical community seems positive on ngVLA.  For example, 83 science white papers were submitted for Astro2020 that mentioned ngVLA specifically by name, more than any other new ground-based facility under consideration.  In addition, the project itself has submitted an APC white paper to the Astro2020 panel for their review, and plans to provide a more comprehensive presentation to the panel itself in late 2019.  In preparation, the project has produced extensive detailed documentation, including the aforementioned {\it ngVLA Reference Design} and cost estimates, to demonstrate its readiness.  

The US ELT program, which is seeking an improved ranking over its Astro2010 standing to secure additional NSF construction funds, could be viewed as a potential competitor to ngVLA in the category of major ground-based facilities.  The construction schedules for the US ELTs and the ngVLA are actually quite complementary, however, with the former needing funds in the early 2020s and the latter in the later 2020s/early 2030s.  For this reason, the two projects do not consider themselves to be in competition.  The success of either project, however, will require an increase in NSF-MREFC astronomy funding by a factor of $\sim$3 over the 2020s.  For context, U.S. Congressional support of NSF as a whole has been steadily increasing over the past few years.

Another risk to ngVLA funding is the possibility that international participation will be far smaller than the 25\% goal.  In response, ngVLA has been widely promoted internationally by NRAO to maximize the project’s visibility and emphasize NRAO’s openness to future partnerships.  For example, an ngVLA-themed science conference was held in Japan in September 2019. As noted earlier, six countries besides the U.S.A. have expressed interest in ngVLA so far, and it is possible more will join when construction plans solidify.

Potential options for ngVLA descope in the event of partial NSF or international funding have not been publicly articulated.  As with ALMA, lower funding levels could be mitigated somewhat by reduced numbers of antennas or receivers, or other initial capability reductions.  In the face of such reductions, however, the impact on science would need to be carefully analyzed.

\subsection{Technological Availability}

Technological availability is a major risk to any cutting-edge scientific endeavor.  To mitigate that risk, the project has minimized exposure to untested and uncertain hardware or software in the {\it ngVLA Reference Design}.  The project aims to build an observatory with components at a high degree of technical readiness, i.e., one that could be completed entirely with today’s technology.  Indeed, this approach has contributed to the maturity of the {\it ngVLA Reference Design} and consequently the robustness of the detailed ngVLA cost estimates.  Nevertheless, ongoing research and development in new technologies for ngVLA will continue, with the goals of improving performance and reducing operations costs.  These initiatives include further improvements to NRC’s single composite reflector design, as well as a custom ASIC serializer-digitizer.  On the software side, NRAO has already invested significant costs to systems currently in use at both ALMA and the VLA, which will likely continue to evolve in performance and capability in the coming decades.

Here it also bears mentioning NRAO’s exemplary track record in constructing facilities.  For example, the VLA, the VLBA, and Expanded VLA projects were each completed on time and within budget, and are widely considered to be success stories in observatory construction.

\subsection{SKA}

A peculiarly Canadian risk for the ngVLA is the perception that the SKA is the single answer to Canada’s future radio astronomy needs.  Granted, the SKA as a concept has been around for a considerably longer time than ngVLA, and was in fact discussed in the very first LRP report in 2000.  Indeed, SKA and ngVLA will overlap at some nominal frequency range, i.e., $\sim$1.2-15 GHz, but ngVLA’s core science goals differ significantly from those of the SKA.  At those frequencies in common, the northern ngVLA and southern SKA together will allow coverage of the entire sky.
Rather than being viewed as being in competition, the two facilities will complement each other, providing Canadians access to next generation radio observing capabilities almost continuously from 70 MHz to 115 GHz (or to 950 GHz, including ALMA).  This enviable situation positions Canada to be a major player in global radio astronomy in the 2030s.

An associated risk is the perception that while SKA requires concrete national participation for access, Canada will get access to ngVLA ``for free” under the generous NSF policy of ``open skies.”  Canada has benefited greatly from this policy, particularly with the VLA.  Though Canada did join with the U.S.A. to form the North American Partnership for Radio Astronomy (NAPRA) as part of its engagement into ALMA, a policy that would have preserved Canadian access to NRAO facilities if “open skies” had changed, that agreement formally expired in 2017.  (Our partnership with NRAO for ALMA is governed by a separate MoU which remains in force.)  More recently, however, the NSF has begun to shift to a less asymmetrical view of “open skies” and now prefers allowing access under fairer reciprocity arrangements.  As a result, 100\% access to ngVLA via “open skies” cannot be assumed going forward.  For context, the SKA plans to provide 5\% of its time to non-member projects.   If SKA does not provide wider access to the U.S. community, the U.S. could significantly restrict ngVLA access to all SKA member countries, including Canada if it joins SKA.

Fortunately, the ngVLA and SKA projects have actually begun formal discussions in June 2019 to allow greater access to each other’s facilities.  For example, in one possible model SKA member countries could have access of up to 40\% of ngVLA time.  If Canada were not to join ngVLA, however, it would only get access to ngVLA via that block.  Assuming the very unlikely scenario where SKA countries receive all that time, however, given the size of Canada’s SKA participation it would presumably only receive about $\sim$6\% of that share, or a maximum of $\sim$2.5\% of total ngVLA time.  That level of access is far lower than current paper production suggests our engagement in VLA really is.  In addition, by not joining ngVLA, Canada will lose the future opportunity to contribute its cutting-edge technologies in its construction and operations.

\section{Project Schedule}

In addition to integrated costs, Table 1 provides the years expected for the major phases of ngVLA construction and operations.  The Table draws from NRAO's extensive and detailed {\it ngVLA Integrated and Master Schedule}.  The project will apply for significant funding in FY2021 from the NSF’s MREFC program to begin construction.  The Design Phase is expected to begin in 2020 and extend until 2024.  Construction will begin in earnest in 2025 and take about nine years until completion.  As with ALMA, there is certainly the potential for commissioning, science verification, and early science with an incomplete but growing array possibly as early as 2028.

\section{HQP and EDI}

The ngVLA is certain to be a boon in the training of Highly Qualified Personnel (HQP) in the years to come.  The VLA has prided itself on being the source of over 500 PhD dissertations since it began operations in 1980, amid $\sim$12,000 papers with over 500,000 citations combined.  Like its forebear was at the time, the ngVLA will be a transformational observatory and the go-to facility for access to the sky across its defined frequency range.  Science aside, the ngVLA will also present numerous opportunities for the training of engineering HQP, given the significant technical contributions Canada could make to the project.

The ngVLA will also be at the forefront of improving equity, diversity, and inclusivity (EDI) within its staff and userbase.  Both NRC and NRAO are committed to improving participation by women and other underrepresented groups among their staffs in Canada and the U.S. respectively.  Both organizations have opportunities for undergraduate students of any background to spend work terms being directly mentored by the staff. Indeed, there is already mindfulness for diversity representation within the project.  For example, organizational leadership of the June 2018 ngVLA conference in Oregon was shared by three women, one of whom is Canadian.  In addition, over half of the oral presentations at the June 2019 ngVLA conference in Virginia were given by women.  

\section{Summary and Recommendation}
The ngVLA represents a new and exciting opportunity for Canada to demonstrate its scientific and technical leadership in a global project that will be the premier instrument of its kind at $\sim$1.2-115 GHz.  We propose Canada engage in ngVLA at a level consistent with its obligations to ALMA and SKA.  For example, a 7\% participation level translates into US\$150M in total for construction and US\$6.5M/yr for operations.    For further information on ngVLA, see \cite{McKinnon19}.

~




\begin{lrptextbox}[How does the proposed initiative result in fundamental or transformational advances in our understanding of the Universe?]
The high sensitivities and resolutions of the ngVLA will enable transformational views of our universe at $\sim$1.2-115 GHz.  Key Science Goals include: i) probing the terrestrial zone of protoplanetery circumstellar disks; ii) detecting interstellar  prebiotic molecules; iii) tracking the evolution of gas in galaxies across cosmic time; iv) testing General Relativity in new regimes using Galactic Centre pulsars; and v) pinpointing the growth of black holes in the era of multi-messenger astronomy.  The ngVLA will be a versatile facility that will enable fundamental advances on many other astrophysical topics of interest to Canadians.  
\end{lrptextbox}

\begin{lrptextbox}[What are the main scientific risks and how will they be mitigated?]
The major scientific risk for ngVLA is that it will not meet some of its Key Science Goals.  Having a diverse set of KSGs offsets this risk, however, by increasing its likelihood of transformational science in some areas. 
\end{lrptextbox}

\begin{lrptextbox}[Is there the expectation of and capacity for Canadian scientific, technical or strategic leadership?] 
Absolutely.  The ngVLA already involves Canadians in key roles on its Science and Technical Advisory Councils.  Also, the Canadian antenna design is the baseline for the project's perfomance and costing models.  Other Canadian technologies (e.g., receivers, correlators, and data archives) can also make important contributions to ngVLA's construction.  With partnerships in both ngVLA, SKA and ALMA, Canada will be strategically placed as a global leader in radio astronomy.
\end{lrptextbox}

\begin{lrptextbox}[Is there support from, involvement from, and coordination within the relevant Canadian community and more broadly?] 
Yes.  Canadian participation in ngVLA has been well received by the Canadian community who have previously been major users of the VLA (and GBT).  Project news has been disseminated via presentations at Canadian Radio Futures meetings, participation in annual conferences, and the ngVLA-Canada mailing list.
\end{lrptextbox}

\begin{lrptextbox}[Will this program position Canadian astronomy for future opportunities and returns in 2020-2030 or beyond 2030?] 
Yes.  Canada's participation in ngVLA will help it enable future opportunities realizing the expanded capabilities of all major radio observatories, including ALMA2030 and SKA2. Moreover, ngVLA will be a key instrument for probing the radio counterparts to phenomena traced by JWST, ALMA, and the ELTs and the electromagnetic counterparts of gravitational wave events.  
\end{lrptextbox}

\begin{lrptextbox}[In what ways is the cost-benefit ratio, including existing investments and future operating costs, favourable?] 
A relatively modest (6-10\%) contribution to ngVLA would enable Canada's access to a US\$2B transformational facility that will dominate astronomy at $\sim$1.2-115 GHz for years to come.  It will also leverage our domestic strengths and maintain Canada as a global leader in radio astronomy hardware, data, and science. 
\end{lrptextbox}

\begin{lrptextbox}[What are the main programmatic risks
and how will they be mitigated?] 
The ngVLA must obtain significant support from both the U.S. astronomy community and the National Science Foundation to be realized.  These entities have been well engaged by the project, and have helped enable its development through planning and interim funding.  The ngVLA must also obtain international collaborators and has accordingly encouraged partnerships with several nations.  The project has embraced low risk technology to reduce uncertainties and costs.  Finally, perceptions must be offset that SKA is the only answer to Canada's radio needs and that Canadian access to ngVLA can occur via NRAO's ``open skies" policy. This risk is mitigated by the project having continued engagement with the Canadian and larger SKA communities.
\end{lrptextbox}

\begin{lrptextbox}[Does the proposed initiative offer specific tangible benefits to Canadians, including but not limited to interdisciplinary research, industry opportunities, HQP training,
EDI,
outreach or education?] 
The ngVLA will enable opportunities for Canadian industry to gain from the technologies needed for its design and construction, improving Canada's economy and technological acumen.  The ngVLA will be the premier instrument for data at $\sim$1.2-115 GHz for comparison with the next generations of optical/IR telescopes, ALMA, and the SKA. Crossing disciplines, the ngVLA will provide key data for exploring the origins of life beyond the Earth (chemistry and biology) and extremely relativistic environments (physics).  Through its construction, ngVLA will provide opportunities for significant scientific and technological HQP training in equitable, diverse, and inclusive workplaces. Finally, NRAO has historically been very proactive about outreach, including staffing the VLA's Visitor Centre, providing tours, and other educational initiatives, and these are expected to continue in the ngVLA era.  An ngVLA station at DRAO will also be a Canadian avenue for outreach.  For more information about ngVLA, including relevant documents, please visit https://ngvla.nrao.edu.
\end{lrptextbox}

\bibliography{example} 

\begin{thebibliography}{}
\expandafter\ifx\csname natexlab\endcsname\relax\def\natexlab#1{#1}\fi

\bibitem[{{McKinnon} {et~al.}(2019){McKinnon}, {Beasley}, {Murphy}, {Selina},
  {Farnsworth}, \& {Walter}}]{McKinnon19}
{McKinnon}, M., {Beasley}, A., {Murphy}, E., {et~al.} 2019, in \baas, Vol.~51,
  81

\end{thebibliography}

\end{document}